\documentstyle[aps,prl]{revtex} 

\begin{document}
\author{Lutfi Ozyuzer$^{1,2,3}$,
John F.  Zasadzinski$^{1,2}$, Chris Kendziora$^{4}$ 
and Kenneth E.  Gray$^{1}$} 
\title{Tunneling Spectroscopy of Overdoped \\
Bi$_{2}$Sr$_{2}$CaCu$_{2}$O$_{8+\delta}$ Single Crystals}

\address{$^{1}$ Science and Technology Center for Superconductivity\\
	and Materials Science Division, Argonne National Laboratory,
	Argonne IL 60439}
\address{$^{2}$ Illinois Institute of Technology, Chicago, IL 60616}
\address{$^{3}$ Department of Physics, 
Izmir Institute of Technology, TR-35230 Izmir, Turkey}
\address{$^{4}$ Naval Research Laboratory, Washington DC 20375}

\maketitle

\begin{abstract}

The point contact tunneling technique is used to examine quasiparticle 
and Josephson currents in overdoped Bi$_{2}$Sr$_{2}$CaCu$_{2}$O$_{8+\delta}$ 
(Bi-2212) single 
crystals with bulk T$_{c}$ values near 62 K.  High quality 
superconductor-insulator-normal metal (SIN) tunnel junctions 
are formed between Bi-2212 crystals and a Au tip, which 
display well-resolved quasiparticle gap features including 
sharp conductance peaks.  Reproducible superconductor-
insulator-superconductor (SIS) tunnel junctions are also 
obtained between two pieces of the Bi-2212 crystals, 
resulting in simultaneous quasiparticle and Josephson currents.  
Superconducting gap values are obtained from the SIS data using a 
model with d-wave symmetry.  The SIS tunneling conductances 
also show very pronounced symmetric dips and humps at high 
bias voltages.  The high bias conductance data reveal that 
the dip and hump features are part of a larger spectrum that 
extends out to 300-400 mV.  The dynamic conductances of both 
SIN and SIS junctions are qualitatively similar to those found 
on optimally-doped Bi-2212 but with reduced gap values 
($\Delta$=15-20 meV).  The maximum Josephson current depends on junction 
resistance in a manner consistent with Ambegaokar-Baratoff theory but 
with a reduced $I_{c}R_{n}$ product of $\sim$2.4 mV.

\end{abstract}

\section{INTRODUCTION}

Many of the early studies on the high-T$_{c}$ superconductors (HTS) 
focused on the physical and chemical properties of 
optimally-doped compounds where the critical temperature, 
T$_{c}$, is the highest value.  Recently, the doping dependence 
of the HTS has become a significant issue due to its 
peculiar influence on superconducting properties such  
the energy gap1 and spectral features of the tunneling 
density of states (DOS).\cite{1,2} One question that arises is how the 
pairing symmetry is affected by doping.  The symmetry of the 
superconducting order parameter for the optimally-doped HTS is 
consistent with $d_{x^{2}-y^{2}}$ (d-wave) as found in angle resolved photo 
emission spectroscopy (ARPES) \cite{3,4} experiments on 
Bi$_{2}$Sr$_{2}$CaCu$_{2}$O$_{8+\delta}$ (Bi-2212) and tricrystal 
ring\cite{5,6} experiments on YBa$_{2}$Cu$_{3}$O$_{7-x}$ and 
Tl$_{2}$Ba$_{2}$CuO$_{6+\delta}$.  The possibility of gap symmetry 
conversion from d-wave to s-wave with overdoping of Bi-2212 is 
suggested by Kendziora et al.\cite{7} and Kelley et al.\cite{8} using 
Raman and ARPES studies respectively.  Tunneling measurements of the 
strong coupling ratio, $2\Delta/k_{B}T_{c}$, in Bi-2212 indicate a 
decrease from a value of 9.3 for optimally doped samples to values 
approaching the BCS mean field value with overdoping.\cite{1} It thus has 
become clear that the overdoped region of HTS is interesting in its 
own right and in this article we focus on tunneling spectroscopy 
measurements on overdoped crystals of Bi-2212.  The principal result 
is that the tunneling spectra are qualitatively similar to those found 
on optimally-doped crystals, but with a reduced energy scale, i.e.  
typical $\Delta$ values are 15 meV-20 meV compared to 38 meV for 
optimal doping.\cite{1,2}
 
	The optimally-doped compound of Bi-2212 has a 
T$_{c}$ onset of 95 K and its T$_{c}$ can be reduced either by additional 
oxygen content (overdoping) or by the removal of oxygen with vacuum 
annealing (underdoping).\cite{1,2,9} Thus, Bi-2212 is a perfect 
candidate for studying its physical properties over a large range of 
the doping phase diagram.  Furthermore Bi-2212 can be grown as a 
single crystal with a very low defect density and a very flat surface 
after cleaving making it suitable for surface sensitive experiments 
such as scanning tunneling microscopy (STM), Raman, and ARPES.
 
	Tunneling spectroscopy is one of the most powerful 
methods for studying the DOS near the Fermi level.\cite{10} Measurements 
on planar junctions have revealed detailed information about 
electron-phonon interactions in conventional superconductors.\cite{11} The 
reliability of point contact tunneling (PCT) spectroscopy has been 
proven for conventional superconductors\cite{12} such as Nb where the strong 
coupling phonon structures have been observed similar to that found in 
planar junctions.  PCT and STM generally measure a spatially local 
density of states near the Fermi level on both conventional and HTS.\cite{10} 
Due to the larger contact area of the PCT method, local variations in 
oxygen stoichiometry are expected to have a smaller effect than with 
an STM.  In general, tunneling studies of Bi-2212 by PCT\cite{1,2} and 
STM\cite{2,13,14} have shown that the spectra are very similar.  The short 
coherence length and anisotropic structure of the HTS present 
significant obstacles to obtaining clear tunneling effects.  Tunneling 
requires a pinhole-free, insulating, homogeneous barrier and clean 
electrode surfaces.  The larger junction area of planar junctions 
increases the probability of finding pinholes and/or other defects and 
the resulting junction produces complex and unexplainable spectra in 
many cases.\cite{15} More than a decade has passed since the discovery of HTS 
and there are still no reproducible planar junctions which show the 
same high quality characteristics as observed in PCT or STM.

	Near-optimal doped Bi-2212 has been examined by a 
variety of tunneling methods\cite{10} including break junctions\cite{1,16,17} , 
PCT \cite{1,2,18} and STM\cite{1,2,13,14,19} and there is a wide scatter in the 
measured values of the superconducting gap parameter, $\Delta$.  Reported 
values of $\Delta$ range from 20 meV to 50 meV.\cite{10} Recent 
measurements by STM, PCT and break junctions\cite{1,2} on high quality 
crystals show that this wide spread is likely due to the sensitive 
dependence of the superconducting gap on hole doping which tends to 
exacerbate problems associated with inhomogeneous doping or surface 
effects.  Because of better understanding of HTS chemistry and 
fabrication, the quality of Bi-2212 crystals has improved 
considerably.  Better control of purity of the crystal narrows the 
spectrum of obtained energy gap values for a given doping level and 
now it can be safely stated that for optimally-doped Bi-2212 with a T$_{c}$ 
onset of 95 K, $\Delta$ is close to 38 meV.\cite{2} This improved 
reproducibility has allowed the observation by PCT and break junctions 
of a remarkable effect whereby the superconducting gap in Bi-2212 
changes substantially and monotonically with doping over a narrow 
doping region where the T$_{c}$ has a maximum.\cite{1} On the overdoped 
side this effect has been seen by STM as well.\cite{2,13,14,19}

	In this paper, we report point contact tunneling 
measurements on single crystals of heavily-overdoped 
Bi-2212 single crystals with bulk T$_{c}$ near 62 K.  Some of the results 
have been published in Ref.  1 cited above, but here we present a more 
detailed examination of the spectra.  The tunneling spectra from both 
superconductor-insulator-normal metal (SIN) and 
superconductor-insulator-superconductor (SIS) junctions exhibit the 
same qualitative features as found on optimally-doped crystals but 
with a reduced gap size.  Typical spectral features of SIN junctions 
include a weakly decreasing background conductance, asymmetric 
conductance peaks, and a dip/hump structure at high bias.  The SIS 
junctions exhibit a well-defined Josephson current when the junction 
conductance exceeds a critical value of about 10 $\mu$S and 
measurements of the Josephson $I_{c}$ vs.$R_{n}$ are presented over 
two decades of junction resistance, $R_{n}$.  For two SIS junctions 
which exhibit clean quasiparticle conductances a comparison is made 
between two model fits to the data: a smeared BCS DOS and a 
momentum-averaged d-wave DOS.  We find that the d-wave fit is in 
better agreement with the sub-gap region but that neither model 
adequately fits the data beyond 2$\Delta$.

\section{EXPERIMENT}

	Samples of single crystal Bi-2212 were grown by a self-flux technique in 
a strong thermal gradient to stabilize the direction of solidification.  
Overdoping has been accomplished using stainless steel cells sealed with 
samples immersed in liquid oxygen, as described elsewhere.\cite{9}  Tunneling 
measurements were performed on crystals with T$_{c}$ nearly 62 K, and a 
transition width of about 1 K.  Magnetization measurements were taken 
after each tunneling measurement, and no significant change of T$_{c}$ has 
been observed.  Tunneling measurements were done with the apparatus\cite{20} 
cooled by $^{4}$He exchange gas coupled to a liquid helium bath.  Cleaved 
single crystal samples of overdoped Bi-2212 usually have shiny surface 
on the a-b plane.  Each is mounted on a substrate using an epoxy so 
that the tip approaches along the c-axis.  Normally a Au tip was used 
as a counter-electrode, and it was mechanically and chemically cleaned 
before each run.

	The SIN and SIS junctions are formed as follows.  After the sample is 
placed in the measurement system and cooled down to 4.2 K, the distance 
and contact force between the tip and sample are adjusted using a 
differential micrometer.  In the ordinary way, a tip pushes onto the 
surface of the crystal and a stable mechanical junction forms between the 
tip and crystal.  Here the insulating tunnel barrier is the native surface 
layer of the crystal.  The two neighboring Bi-O planes in Bi-2212 have 
weak bonds compared to Cu-O and Sr-O layers.  Therefore the cleaved 
surface of Bi-2212 is a single Bi-O plane and presumably the barrier is 
due to this layer and the underlying Sr-O layer.  STM studies have 
indicated that this surface layer is semiconducting with an energy gap of 
at least 100 meV.\cite{15}  While the tip is pushed against the crystal, the I-V 
curve is continuously monitored on an oscilloscope until a suitable 
junction is obtained.  Such junctions clearly display a superconducting 
gap in the I-V characteristics.  The junction resistances range from 300 $\Omega$ 
to 50 k$\Omega$ for the SIN junctions but the majority of junctions fall in a 
narrower range $\sim$ 2k $\Omega$-10 k$\Omega$.  Although the tip approaches 
nominally along the c-axis the tunneling direction is not known with 
certainty, but we will argue later that both the SIN and SIS junctions 
have the tunnel barrier perpendicular to the c-axis.  The junction 
area is also not known with certainty, however, a good estimate is 
found from similar PCT measurements of Nb\cite{12} where the barrier height 
analysis of the tunneling conductance led to a contact diameter of 
$\sim$2400 $\AA$.

	Increasing the force of the tip against the crystal produces an ohmic 
contact ($\sim$1 $\Omega$) between the tip and crystal, likely due to a 
perforation of the barrier layer.  This results in a mechanical bond 
between Au and Bi-2212 because when the tip is elevated, often a piece 
of Bi-2212 crystal also goes up that is fastened to the tip.  We have 
found that it is not necessary to raise the tip to form an SIS 
junction but merely to relieve the pressure.  Presumably in this case 
one of the Bi-O plane pairs are separated and this results in an SIS 
junction, which forms between different parts of the crystal.  The 
dislodged piece probably remains in the cavity in the larger crystal 
as these SIS junctions are mechanically very stable.  As a consequence 
this technique provides a fresh junction that is formed in-situ deep 
in the crystal, thereby minimizing the exposure of the surface.  A 
schematic of the SIS junction between two pieces of crystal is 
represented in Fig.  1.  The junction area is uncertain, however, for 
a rough idea, the size of the attached crystallite is on the order of 
10-100 $\mu$m on edge so this provides an upper bound to the area.  
SIS junction resistances are again in the range, 2 k$\Omega$ to 120 
k$\Omega$ so this indicates that the actual contact area is 
considerably less than 10 $\mu$m on edge.  The first derivative 
measurements, $\sigma$=dI/dV, were obtained using a Kelvin bridge 
circuit with the usual lock-in procedure.  I(V) and dI/dV(V) were 
simultaneously plotted on a chart and recorded on computer.

	  The SIN junctions and SIS junctions were easily identified by their 
characteristic tunneling conductances.  SIN junctions display an 
asymmetric, weakly decreasing background, an asymmetric dip/hump feature 
and as low as 10\% - 15\% zero bias conductance.  The  SIS junctions have a 
symmetric background, less than 1\% zero bias conductance,  symmetric dip 
and hump features and conductance peaks at 2$\Delta$.  In addition the SIS 
junctions exhibit a hysteretic Josephson current when the junction 
resistance is below 100 k$\Omega$.

\section{RESULTS OF SIN JUNCTIONS}

	Before showing the experimental results, we review here the basic 
theoretical approach to analyzing the tunneling data. The tunneling 
current in an SIN junction can written as,\cite{11}

\begin{equation}
I(V)=c\int |T|^{2}N_{sn}(E)N_{nn}(E+eV)[f(E)-f(E+eV)]dE
\end{equation}
where $f(E)=[1+exp(E/k_{B}T)]^{-1}$ is the Fermi function, $N_{sn}(E)$ 
is the DOS of the superconductor, $N_{nn}(E)$ is DOS of the normal metal 
which is Au in our case, $|T|^{2}$ is the tunneling matrix element, and c 
is a proportionality constant.  The quasiparticle energy is given by 
$E$, which is defined relative to the Fermi level.  For a rough 
estimation of the tunneling current, we may assume that $N_{nn}$ is a 
constant near Fermi level and tunneling matrix element, $|T|^{2}$, has weak 
energy dependence over the voltage range of the energy gap.  Then,

\begin{equation}
 dI/dV\equiv \sigma _{s}=c|T|^{2}N_{sn}(E)=c|T|^{2}N_{s}(E)N_{n}(E)
\end{equation}
where we have now set $E=eV$.  $N_{s}(E)$ is the superconducting 
part of the DOS, and $N_{n}(E)$ is normal state DOS of superconductor.  
In most low-T$_{c}$ superconductors, $N_{n}(E)$ usually is a constant over 
the energy range of interest.  The ratio of $\sigma_{s}/\sigma_{n}$ 
gives the superconducting DOS.  According to BCS theory, the 
superconducting DOS is

\begin{equation}
N_{s}(E) = \left\{ \begin{array}{ll} Re[\frac {E}{\sqrt{E^2-\Delta ^2}}] & 
\mbox{$|E|\geq \Delta$} \\
                 0                                 & \mbox{$|E|<\Delta$}
                 \end{array}
           \right. 
\end{equation}

For conventional superconductors, the energy gap, 
$\Delta$, is a constant in ${\bf k}$ space (isotropic s-wave pairing 
symmetry).  However, for superconductors with d-wave pairing symmetry, 
the energy gap is anisotropic on the Fermi surface.  In a two 
dimensional superconductor, a simple d-wave DOS suggested by Won and 
Maki\cite{21} is given by

\begin{equation}
N_s(E,{\bf k})=
\mathop{\rm Re}
\left\{ \frac{E-i\Gamma }{\sqrt{(E-i\Gamma )^2-\Delta ({\bf k})^2}}\right\} 
\end{equation}
where $\Delta({\bf k})=\Delta_{0}$cos$(2\phi)$ is the k dependent energy gap 
and $\phi$ is the polar angle in k-space.  The DOS, $N_{s}(E)$, is 
found by an integral over the polar angle $\phi$.  Here, $E$ is replaced 
by $E-i\Gamma$, where $\Gamma$ is a smearing parameter to account for 
quasiparticle lifetime.\cite{22}

	Figure 2  shows the dynamic conductances of three SIN junctions at 4.2 K 
(dots) obtained from three different crystals.  Negative voltage 
corresponds to removal of electrons from the superconductor, or occupied 
quasiparticle states in the DOS.  Each junction displays sharp conductance 
peaks near 20 mV and a decreasing background conductance out to 200 mV.  
Figure 2(a) and 2(b) display a very pronounced dip and hump for negative 
bias that is also seen in tunneling studies\cite{1,2,13,14} of optimally-doped 
Bi-2212 and is consistent with ARPES\cite{3,4} measurements of the spectral 
weight function along the ($\pi$,0) momentum direction.  This dip feature is 
located at approximately twice the voltage of the conductance peak and 
hence is scaling with the superconducting gap itself.  A possible 
explanation for the origin of the dip feature is given in Ref. 2.  The 
zero bias conductances vary from 20\% to 40\% of the estimated normal state 
conductance but in a few cases we have seen values as low as 10\%-15\% in 
overdoped crystals. 
 
	The tunneling conductance in Fig. 2(c) is deliberately chosen to 
demonstrate the general features of a junction with a relatively low 
resistance, about 300 $\Omega$ in this case.  Such a low resistance 
means that we are approaching the point contact situation where 
Andreev reflection\cite{23} may be taking place and this might explain the 
more sharply increasing conductance as zero bias is approached.  
Kashiwaya et al.\cite{24} calculated the differential conductance of a normal 
metal-insulator-(d-wave) superconductor junction within the framework 
of the BTK theory\cite{25} which includes tunneling and Andreev reflection.  
Further study is needed to fit Fig.  2(c) and other low resistance 
junctions to the theory, which is beyond the scope of this paper.  
However we note that a general feature of this model\cite{24} is the 
presence of a sharp zero-bias conductance peak in the superconducting 
state when the tunnel barrier is perpendicular to the Cu-O planes of 
HTS.  Since we never observe this feature, this seems to suggest that 
our SIN junctions all have the barrier parallel to the Cu-O planes, as 
would result if it originates from the native Bi-O layer on the 
cleaved crystal surface.
   
	Because of large upper critical fields\cite{26} of HTS, the normal state 
conductance cannot be obtained experimentally.  Somehow we have to 
estimate normal state conductance.  Since we expect the superconducting 
and normal state conductances to merge at voltages much higher than the 
gap, we can extrapolate the values of $\sigma_{s}$ at high voltages 
and get an approximation, $\sigma_{n}*$, thus obtaining an approximate DOS from 
the ratio of $\sigma_{s}$/$\sigma_{n}*$.  We used a simple polynomial fit to 
obtain $\sigma_{n}*$ and checked the resulting normalized conductance 
to make sure that the sum rule of the DOS was verified.  The solid 
lines in the Fig.  2 represent estimated normal state conductance 
curve using this procedure.

When the superconducting dynamic conductance is divided by $\sigma_{n}*$, 
we obtain the normalized conductances [dots in Fig.  3], which can be 
compared to theory to find the energy gap of the superconductor.  Note 
that the dip features still exist after normalization and in the case 
of Fig.  3(c), the dips are now evident whereas they were obscured in 
the raw conductance data.  In Fig.  3, solid lines correspond to fits 
from eq.  (1) using a smeared BCS DOS i.e.  Eq.  (3) with the 
inclusion of a quasiparticle lifetime term, $\Gamma$.  Energy gaps 
vary from 15 to 20 mV for the SIN junctions.  Figure 3(a) shows a good 
overall agreement with the smeared BCS DOS while Fig.  3(b) and Fig.  
3(c) show a sub-gap conductance that is more cusp-like and suggestive 
of d-wave symmetry.  Fits of the data in Fig.  3(b) to a d-wave DOS as 
in eq.  (4) lead to an identical value for $\Delta$ (19 meV) which is 
then interpreted as the maximum of $\Delta$(${\bf k}$).  This range of sub-gap 
shape from d-wave-like to smeared s-wave-like was also observed\cite{2} in 
optimally-doped Bi-2212 and one possible explanation has to do with 
tunneling directionality effects.\cite{27} For optimally-doped 
samples, reproducible energy gaps are found\cite{1,2} near 38 meV 
corresponding to $2\Delta/k_{B}T_{c}$ $\sim$ 9.3.  This is very large when 
compared to a BCS ratio of 4.28 for a d-wave superconductor.\cite{21} 
In the overdoped samples, we obtained 2$\Delta$/k$_{B}$T$_{c}$ in the range 
of 5.6-7.5.  The energy gap decreases rapidly when T$_{c}$ goes from 
95 to 62 K and the trend is to approach the BCS mean-field gap value.  
This rapid decrease of the gap with increased hole doping is also seen 
in ARPES experiments\cite{28} and STM measurements\cite{2,14} on 
overdoped Bi-2212.

\section{RESULTS OF SIS JUNCTIONS}

	If both electrodes are identical superconductors in eq. (1) the 
quasiparticle conductance peaks are at $\pm$2$\Delta$.  As a result of the 
convolution of the superconducting DOS, the quasiparticle peaks are 
sharper and zero bias conductances are lower than SIN junctions.  The 
Fermi function effect is very small compared to SIN junctions.  In 
addition, Cooper pairs may tunnel through the barrier and a Josephson 
current occurs at zero bias in the I-V characteristics.  

	In this part of the paper, highly reproducible and stable SIS junctions 
will be examined.  Figure 4(a), (b) and (c) show current-voltage 
characteristics of three junctions from three different crystals.  These 
were measured at 4.2 K and the junctions indicate SIS type, with very low 
leakage currents and sharp current onsets at eV$\sim$40 meV which is twice the 
gap value found in the SIN junctions.  They also exhibit a hysteretic 
Josephson current at zero bias.  Figure 4(d) shows the sub-gap region of 
Fig. 4(c) on a more sensitive scale to clearly show the hysteretic nature 
of the Josephson current.
   
	Figure 5 shows conductance-voltage characteristics of SIS junctions from 
three different samples at 4.2 K.  The junction resistances are in the 
range 5 k$\Omega$ - 15 k$\Omega$.  Here the Josephson current gives rise to a 
very large conductance peak at zero bias which is off the scale.  
Peaks correspond to $\pm$2$\Delta$ values.  Symmetric dips are found 
near 3$\Delta$ as found on Bi-2212 crystals near optimal doping.\cite{1,29} On 
average, the dip strength is low with respect to optimally-doped 
samples.\cite{1,2} The conductance at higher bias voltages ($e|V|$ $>$ 
3$\Delta$) at first decreases with increasing bias, as found in the 
SIN junctions.  This behavior exists out to $|V|$ $\sim$ 300 mV but 
for bias voltages beyond this value there is an upturn in the 
conductance.  A similar upturn is also observed in break junctions of 
Bi-2212 by Mandrus et al.\cite{16} In the simple approximation 
expressed in eq.(2), it is assumed that the tunneling matrix element, 
$|T|^{2}$, is energy independent.  However, at very high bias voltages 
we are likely approaching the barrier height potential.  As a 
consequence of this, the tunneling transmission probability is 
increasing and the barrier height effect which gives a parabolic 
increase in conductance\cite{11} may become dominant beyond $\pm$300 mV.
    
	The fact that the upturn in conductance occurs at nearly the same bias 
value in these junctions as well as other break junctions\cite{16} which can 
differ in resistance by two orders of magnitude strongly argues against 
extrinsic effects such as heating as the cause of the initial decreasing 
background between 100 mV and 300 mV.  Rather, the data suggest that the 
dip and hump feature are part of a larger spectrum that includes a tail 
extending out to 300-400 mV and that all of these features are intrinsic 
effects of the quasiparticle DOS in Bi-2212.  It should be noted that a 
similar dip-hump-tail feature is found in the quasiparticle spectral 
weight function, A({\bf k},$\omega$) measured in ARPES.\cite{30} Although no 
definitive explanation of these spectral features exists, it is 
important to note the consistency between tunneling and ARPES.

	In our preliminary analysis of these SIS data, the goal was to obtain the 
energy gap of the Bi-2212, and we used the simplest approach, which was to 
use Eq. (1) with a smeared BCS DOS for both electrodes.  Furthermore, one 
of the key differences between s-wave and d-wave models of the SIS 
conductance is in the shape of the sub-gap region which is difficult to 
obtain experimentally because of the switching property of the Josephson 
current.  Thus a detailed comparison of the two models for every junction 
is not fruitful and we found that an accurate estimate of 2$\Delta$ could be 
obtained directly from the conductance peak voltage. To demonstrate this, 
the data of Figs. 5(a) and 5(b) were normalized by a constant value which 
is the conductance at 150 mV.  The normalized conductances are displayed 
in dots in  Fig. 6(a) and 6(b).  The steeper background shape in Fig. 5(c) 
did not allow us to normalize it with constant value, so we chose another 
junction (which was not measured to a high bias voltage) for normalization 
in Fig. 6(c).   The solid lines in Fig. 6, represents SIS fits with the 
smeared BCS DOS and T=4.2 K.  The $\Delta$ and $\Gamma$ values are shown 
for each curve.  Note that in each case the conductance peak voltage 
is exactly 2$\Delta$/e.  The quality of the fits in the sub-gap region 
varies but the significance of this is difficult to asses since 
switching of the Josephson current can introduce structural artifacts 
near zero bias.  The principal result here is that the energy gap 
values ($\Delta$= 16 meV - 18 meV) are consistent with those found in 
SIN junctions.  Considering all of the SIS junctions ($>$ 20) on the 
four different crystals, the full range of gap values was $\Delta$ = 
15 meV - 20 meV.

  	Up to now, we used a simple smeared BCS DOS for analysis of the SIN and 
SIS conductance curves to obtain the energy gap of the overdoped Bi-2212.  
SIN and SIS fits gave consistent results and the energy gap was found to 
be dramatically smaller than that of optimally-doped Bi-2212.  However, it 
is clear from the fits in Figures 3 and 6 that a smeared BCS DOS does not 
adequately fit the data in each case.  Given the expected d-wave symmetry 
of the gap in Bi-2212, we analyzed some of the conductance data using the 
simple d-wave model of Eq. 4.  We focus on SIS junctions because it is in 
these types of junctions that the differences between d-wave and smeared 
s-wave are most clearly observed.  Furthermore, we choose relatively high 
resistance junctions where the effects of the Josephson current are 
minimized. 
  
	The inset of Fig.7 shows another SIS tunneling conductance which is 
particularly clean.  The Josephson current was very low for this 
particular junction due to its higher resistance ($\sim$40 k$\Omega$) and a 
relatively small zero bias conductance peak is observed.  To obtain 
the sub-gap conductance very clearly, the d.c.  bias current was swept 
very slowly inside the gap in both directions.  These data also 
display generic Bi-2212 quasiparticle tunneling conductance 
characteristics such as sharp quasiparticle peaks, and a dip/hump 
feature at high bias.  Because the data are symmetric with bias 
polarity, we focus only on the positive bias side of the spectrum.  
Our goal is to examine to what extent the shape of the sub-gap 
conductance reveals information on pairing symmetry.  Thus we consider 
the simple d-wave model of eq.  (4) for comparison to the smeared BCS 
model.
 
  	The solid line in Fig. 7 corresponds to the estimated normal state 
conductance, $\sigma_{n}*$.  It is obtained after masking the data below 
100 mV and fitting with a 3rd order polynomial curve.  The dots in 
Fig.  8 show the normalized tunneling conductance curve from Fig.  7 
and the solid line is the smeared BCS fit, including $\Gamma$, which 
is obtained from Eq.  (1) with T=4.2 K to obtain the SIS tunneling 
conductance.  The value of 2$\Delta$=35 meV is consistent with the 
other junctions and as noted above could have been obtained directly 
from the conductance peak voltage.  The overall quality of the fit is 
poor.  The smaller peak in the BCS fit near $\Delta$ arises from the 
$\Gamma$ parameter and while the data do not show such a distinct 
feature there is clearly a broadened peak in this region.  The 
dip/hump feature is not reproduced in this simple model.

	Before discussing the d-wave fit of the data, it is useful to present the 
generic features of the simplest model.  Figure 9(a) shows a generated DOS 
curve for d-wave symmetry using Eq. (4).  The underlying assumptions are 
that the Fermi surface is isotropic and the probability of tunneling is 
same for all k directions.  The cusp feature at zero bias is the basic 
indication of d-wave pairing symmetry.  The SIS tunneling conductance 
curve is obtained with convolution of Fig. 9(a) with itself using Eq. (1), 
with T=4.2 K, and this is shown in Fig. 9(b).  Note that the peak height 
to background (PHB) ratio is $\sim$ 2 for the model d-wave SIS tunneling 
conductance whereas the normalized SIS curves of Figs. 6 and 8 suggest a 
PHB ratio that is much higher than 2.  Thus we see at the outset that this 
simple d-wave model will not fit our data.  One possible consideration is 
that our estimated normal state curve is incorrect.  While this cannot be 
ruled out, it is difficult to conceive of a normal state conductance that 
would bring the PHB ratio from a value of 4 or 5 down to a value of 2 as 
the d-wave model requires.  
   
	  Another possibility is that one or more of the underlying assumptions 
of the simple model are incorrect.  ARPES results reveal that Bi-2212 has 
arch like Fermi surfaces in the four corners of the Brillouin zone\cite{30} and 
this suggests the possibility that particular regions of momentum space 
are more heavily weighted than others in the contribution to the tunneling 
current, either due directly to the band structure or to the tunneling 
matrix element.  For this reason we include a weighting function, 
$f(\phi)$, to Eq.  (4) as is in Refs.  31 and 32,

\begin{equation}
N_s(E,\phi)= f(\phi) \mathop{\rm Re} \left\{ \frac{E-i\Gamma }{\sqrt{(E-i\Gamma 
)^2-\Delta (\phi)^2}}\right\}
\end{equation}
	
where $f(\phi)=1+\alpha cos(4\phi)$.  and the angle $\phi$ is measured 
with respect to the ($\pi$,0) direction in the Brillouin zone.  Here 
$\alpha$ is a directionality strength and the full quasiparticle DOS 
is obtained by integrating over $\phi$.  It should be emphasized that 
eq.(5) is a phenomenological expression and there is no rigorous 
derivation of this result, but it can be seen immediately that this 
weighting function will qualitatively reproduce some of the 
conductance features observed in the experiment.  For example, the 
observation in some SIN junctions of a smeared BCS DOS shape to the 
gap region conductance may be attributed to a non-zero a value which 
more heavily weights contributions along the ($\pi$,0) direction where 
$\Delta({\bf k})$ is largest and minimizes the contribution from the nodal 
directions.  This weighting would also explain the large PHB ratios in 
SIS junctions since a smeared BCS DOS can exhibit a large PHB as shown 
in Fig.  8.
    
	We fit the normalized SIS tunneling conductance, given in Fig. 8, using 
Eq. (5) and the solid line in Fig. 10 shows the resulting curve with 
$\Delta$=17.5 meV, $\Gamma$=0.05, $\alpha$=0.8.  Note here that $\Delta$ is the maximum of 
$\Delta({\bf k})$ and is the same magnitude as obtained with the BCS fit.  For 
eV$<$2$\Delta$, the experimental results and fit are in better agreement than 
for the BCS fit in Fig.  8.  For eV $>$ 2$\Delta$ the fit shows an abrupt drop 
from the peak to background that is not found in the data.  This 
feature appears to be generic to d-wave models (see Fig.  9(b)) but we 
have never observed it in over 50 SIS junctions on Bi-2212 with 
various hole doping\cite{1,2} and to our knowledge such an abrupt drop has 
never been seen in the literature.  Rather the experimental data 
invariably exhibit a much broader conductance peak followed by the 
dip/hump feature which is not found in this simple d-wave model.  The 
inset of Fig.  11 shows another SIS junction which was chosen for 
analysis because of the relatively low Josephson current and clean 
quasiparticle data.  This junction exhibits a smaller PHB ratio than 
in Fig.  10.  The solid line in the Fig.  11 is the estimated normal 
state conductance.  The normalized data is displayed in Fig.  12 along 
with the fit obtained with the following parameters: $\Delta$=19.5 
meV, $\Gamma$=1.5, $\alpha$=0.6.  The larger $\Gamma$ value reflects 
the generally broader features of this junction's spectrum and there 
is better agreement with data in the sub-gap region.  Nevertheless, 
for eV$>$2$\Delta$ the data display the same deviation from the fit as 
found in Fig.  10.  This suggests the possibility that these 
characteristics of the SIS spectra for eV$>$2$\Delta$ are intrinsic and 
not due to a less interesting effect such as a distribution of gap 
values due to sample inhomogeneity.  It was recently argued that the 
dip feature is a strong coupling effect\cite{2} analogous to the 
phonon structures observed in conventional strong-coupled 
superconductors.\cite{11} Perhaps the entire spectrum beyond 2$\Delta$ 
requires a full microscopic d-wave model, analogous to strong-coupling 
theory for conventional superconductors.

\section{JOSEPHSON CURRENTS}

	In this part of the paper, we will present and discuss Josephson currents 
on overdoped Bi-2212.   Reproducible Josephson tunnel junctions have been 
obtained, indicating both quasiparticle and Josephson tunneling currents 
simultaneously as we displayed in Fig. 4.  Josephson currents are observed 
at 4.2 K when the junction conductance exceeds a minimum value.  For the 
overdoped crystals here the minimum value was about 10 $\mu$S whereas it was 
about 5 $\mu$S for near optimal doped Bi-2212 (Ref. 29).   The hysteretic 
nature of the Josephson current is shown in Fig.  4(d) and the 
$I_{c}R_{n}$ product for this particular junction is 3 mV.  The 
Ambegaokar-Baratoff (A-B) theory\cite{33} for BCS superconductors 
predicts the relation,

\begin{equation}
I_{c}R_{n}=\frac{\pi \Delta(0)}{2e}
\end{equation}
where $I_{c}$ is the Josephson current, $R_{n}$ is the resistance of the 
junction, and $\Delta$(0) is the energy gap at 0 K.  Figure 13 shows 
the measured critical current $I_{c}$ as function of the junction 
resistance $R_{n}$ for 13 different SIS junctions formed on the four 
overdoped Bi-2212 crystals.  Note that the junction resistance varies 
by about two decades.  The solid line corresponds to the maximum 
theoretical value expected from A-B theory using $\Delta$=17 meV, the 
average gap found in the quasiparticle data.  The data show that the 
maximum Josephson current depends on junction resistance in a manner 
which is consistent with A-B theory, but with a value of $I_{c}R_{n}$ 
that is reduced from that expected using the average quasiparticle 
gap.  We obtained a maximum of 7 mV for $I_{c}R_{n}$ product which is 
approximately 40\%
of the expected magnitude and the overall average is 2.4 mV.  While the 
average $I_{c}R_{n}$ product is small compared to A-B theory it should be 
remembered that this theory applies to s-wave superconductors and assumes 
that the tunneling matrix element, T$_{{\bf k,k'}}$, is a constant, 
independent of electron momentum.  When this type of incoherent 
tunneling matrix element is applied to d-wave superconductors, one 
obtains $I_{c}R_{n}$=0 to leading order.\cite{34,35} This is a 
consequence of the changes in sign of the d-wave order parameter as 
one sums over momentum.  Thus the $I_{c}R_{n}$ products observed here 
are actually quite large for d-wave superconductors.  The reason for 
these large values is not known but it may be an indication that there 
is a coherent part to the tunneling matrix element,\cite{35} that is, 
${\bf k}$ must equal ${\bf k'}$ for quasiparticle states on either 
side of the junction.  In this case the signs of the order parameters 
are multiplied, i.e.  (+1)$^{2}$ or (-1)$^{2}$ and this leads to a 
non-zero Josephson current for d-wave.  More experimental and 
theoretical work is needed to understand the magnitude of the 
Josephson $I_{c}R_{n}$ products observed here.

\section{SUMMARY AND CONCLUSION}

	We have performed PCT measurements on overdoped Bi-2212 single crystals 
with bulk Tc values near 62 K.  High quality reproducible SIN and SIS 
junctions were obtained and the two junction types gave consistent 
results.  Attempts to fit the SIS spectra to a weighted, momentum-averaged 
d-wave DOS led to identical gap values as found with a smeared BCS DOS but 
with a somewhat improved fit to the sub-gap region of the conductance.  
The magnitude of $\Delta$ = 15 meV-20 meV is reproducibly observed in 
both types of junctions on four different crystals.  This gap 
parameter is a significant reduction from that found on optimally 
doped crystals ($\Delta$$\sim$38 meV).\cite{1,2} The value of the strong 
coupling parameter, 2$\Delta$/k$_{B}$T$_{c}$, is between 5.6 to 7.5, reduced 
from the value of 9.3 found on optimally-doped samples and this 
indicates a trend toward weak coupling, mean-field behavior as hole 
doping is increased.
 
	The dynamic conductance spectra of the SIN and SIS junctions are 
qualitatively similar to those found on optimally-doped\cite{1,2}	and near 
optimal doped crystals\cite{16} with all junctions exhibiting sub-gap 
conductance, sharp conductance peaks and a dip/hump/tail feature.  The 
position of dip is nearly 2$\Delta$ for SIN junctions and 3$\Delta$ for SIS 
junctions similar to what is found for other hole doping 
concentrations including underdoped.\cite{1,2} Thus the energy scale for 
these spectral features seems to be set by the superconducting gap 
value.  Neither the smeared BCS DOS nor the d-wave DOS gives a good 
agreement with the data for eV$>$2$\Delta$.  In particular, we have never 
observed the sharp decrease in conductance for SIS junctions at eV=2$\Delta$ 
which is predicted by d-wave theory.  It appears that the dip, hump 
and tail features, which are reproducibly found in these SIS spectra, 
are intrinsic properties of the quasiparticles in the superconducting 
state.  These features are also found in the quasiparticle spectral 
weight functions measured by ARPES\cite{30} and thus we have demonstrated an 
important correlation between these two measurements.  What is likely 
needed to explain all the spectral features for eV$>$2$\Delta$ is a full 
microscopic theory, analogous to Eliashberg theory for conventional 
superconductors.

	The sub-gap features and the conductance peaks of the SIS junctions were 
studied in the context of pairing symmetry.  A smeared BCS DOS gives a 
poor fit to the data in these regions.  We have also shown that a pure 
d-wave DOS with a constant tunneling matrix element is also not capable of 
fitting these particular tunnel junctions, in particular the large PHB 
ratio.  This has forced us to assume a weighted average of the d-wave DOS 
with the tunnel current weighted more heavily along the ($\pi$,0) momentum 
direction where the d-wave gap is maximum.  This improves the fit to the 
sub-gap region and also provides a rationale for the varying sub-gap 
shapes from junction to junction as being due to different weighting 
functions.  However it is still not understood why particular tunneling 
directions are preferred.
  
	Josephson currents are observed as a robust feature of the SIS junctions 
when the junction conductance exceeds a minimum value of about 10 
$\mu$S.  The maximum critical current, $I_{c}$, scales with junction 
resistance in a manner consistent with A-B theory for s-wave 
superconductors, but with a value of the $I_{c}R_{n}$ product that is 
reduced from that expected using A-B theory and the measured 
quasiparticle gap.  Nevertheless, the average value of $I_{c}R_{n}$ 
(2.4 mV) and the maximum value (7 mV) are much larger than expected 
for the Josephson current between two d-wave superconductors assuming 
incoherent tunneling between the electrodes as is done in A-B theory.  
This implies that more experimental and theoretical work is necessary 
to understand the Josephson current between two HTS electrodes.

\section{ACKNOWLEDGMENTS}

This work was partially supported by U.S. Department of 
Energy, Division of Basic Energy Sciences-Material Sciences 
under contract No. W-31-109-ENG-38, and the National Science 
Foundation, Office of Science and Technology  Centers 
under contract No.  DMR 91-20000.

\newpage

 FIGURE CAPTIONS

	Figure 1. 	Schematic representation of SIS junction formed by a Au tip.
	
	Figure 2. 	Dynamic conductance of three SIN junctions [dots], from three 
different overdoped Bi-2212 (T$_{c}$=62 K) crystals at 4.2 K.  Negative bias 
region corresponds to occupied states in the DOS.  The solid lines are 
estimated normal state conductances ($\sigma_{n}^{*}$) which are generated 
from verifying the sum rule of the DOS.

	Figure 3. 	Normalized conductance of three SIN junctions [dots], which 
are given in Fig. 2.  The solid lines are BCS DOS fit which include 
smearing factor $\Gamma$.  $\Delta$ and $\Gamma$ values are given in the figure for each fit.
   
	Figure 4.	The current-voltage characteristics of three different SIS 
junctions.  (d) is expanded scale of (c) near zero bias for clarity.  (d) 
shows hysteretic zero-bias feature which indicates a Josephson current. 

	Figure 5.	The tunneling conductance of SIS junctions for overdoped 
Bi-2212.  The upturn around $\pm$300-400 mV may indicate that the bias is 
approaching the barrier height.  The sub-gap tunneling conductance can not 
be seen, because of switching.

	Figure 6.	The normalized tunneling conductance of SIS junctions [dots]. 
(a) and (b) are the same data with Fig. 5 (a) and (b).  Figure 6 (c) is a 
different junction.  The normalizations are conducted by a constant value 
which are chosen at 150 mV. The solid lines are BCS fits, which are 
calculated using given $\Delta$ and $\Gamma$ values at graph and T=4.2 K.

	Figure 7.	The positive bias part of the tunneling conductance of SIS 
junction [dots].   Relatively small Josephson current helped to reveal 
sub-gap conductance very clearly.  The solid line is an estimated normal 
state conductance.  The inset shows full spectrum. 

	Figure 8.	The normalized conductance [dots] of the junction which is 
shown in Fig. 7.  The solid line is an SIS fit for BCS superconductor, 
which is calculated for $\Delta$=17.5 meV, $\Gamma$=1.4 meV and T=4.2 K.
 
	Figure 9.	The momentum-averaged DOS for a superconductor with $d_{x^{2}-y^{2}}$  
pairing symmetry (a).  The convolution of (a) with itself produces an SIS 
tunneling conductance (b).  
  
	Figure 10.	The normalized conductance [dots] of the junction which is 
shown in Fig. 7.  The solid line is an SIS fit for a d-wave superconductor 
with weighting, $f(\phi)=1+0.8cos(4\phi)$, which is calculated for 
$\Delta$=17.5 meV, $\Gamma$=0.05 meV and T=4.2 K.
	 
	Figure 11.	The positive bias part of the tunneling conductance of another 
SIS junction [dots].  The solid line is an estimated normal state 
conductance.  The inset shows full spectrum. 

	Figure 12.	The normalized conductance [dots] of the junction which is 
shown in Fig. 11.  The solid line is an SIS fit for a d-wave 
superconductor with weighting, $f(\phi)=1+0.6cos(4\phi)$, which is calculated for 
$\Delta$=19.5 meV, $\Gamma$=1.5 meV and T=4.2 K. 

	Figure 13.	Josephson current versus junction resistance for overdoped 
Bi-2212 at 4.2 K [dots] in logaritmic scale.  The solid line is 
Ambegaokar-Baratoff prediction for $\Delta$=17 meV.

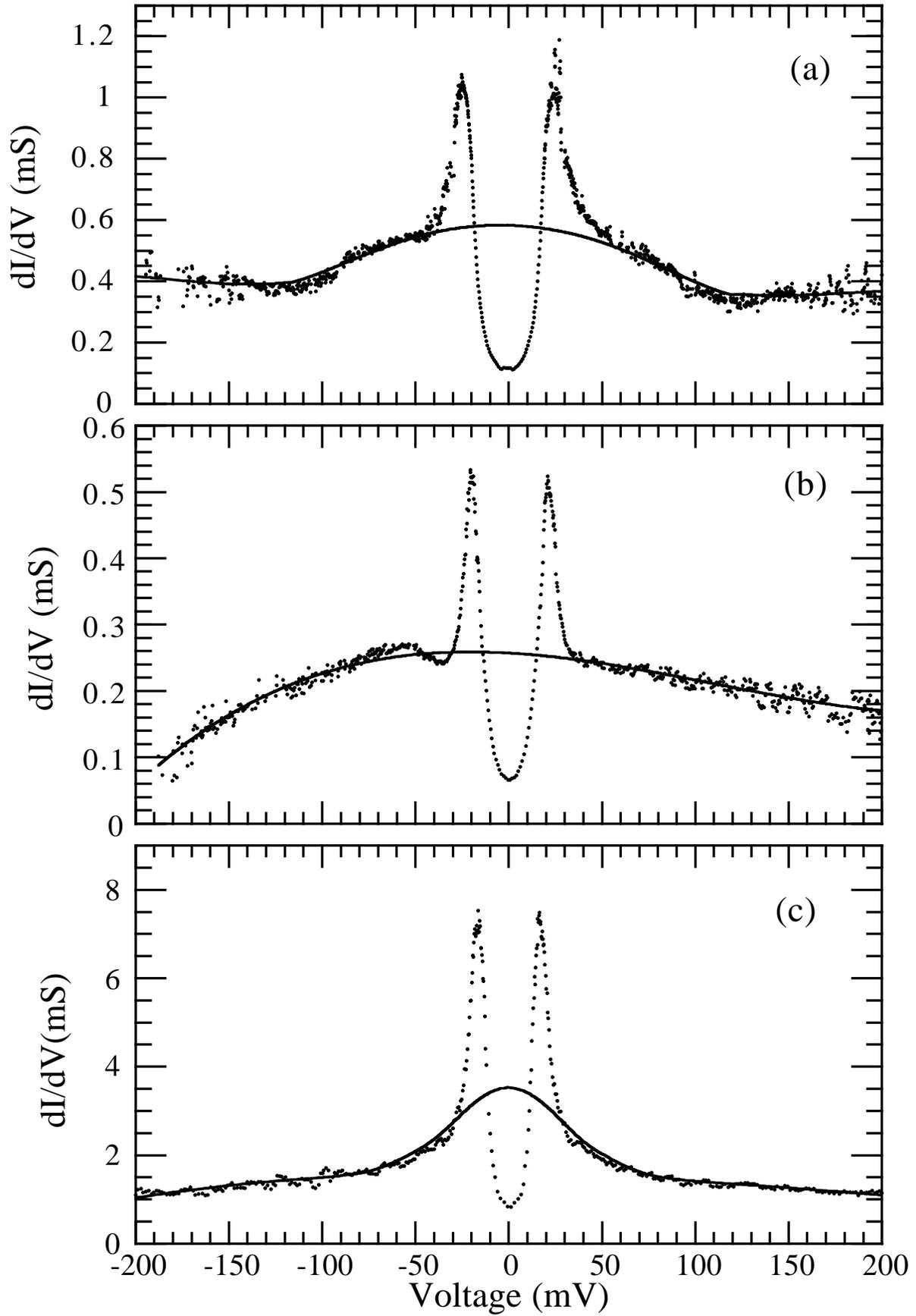
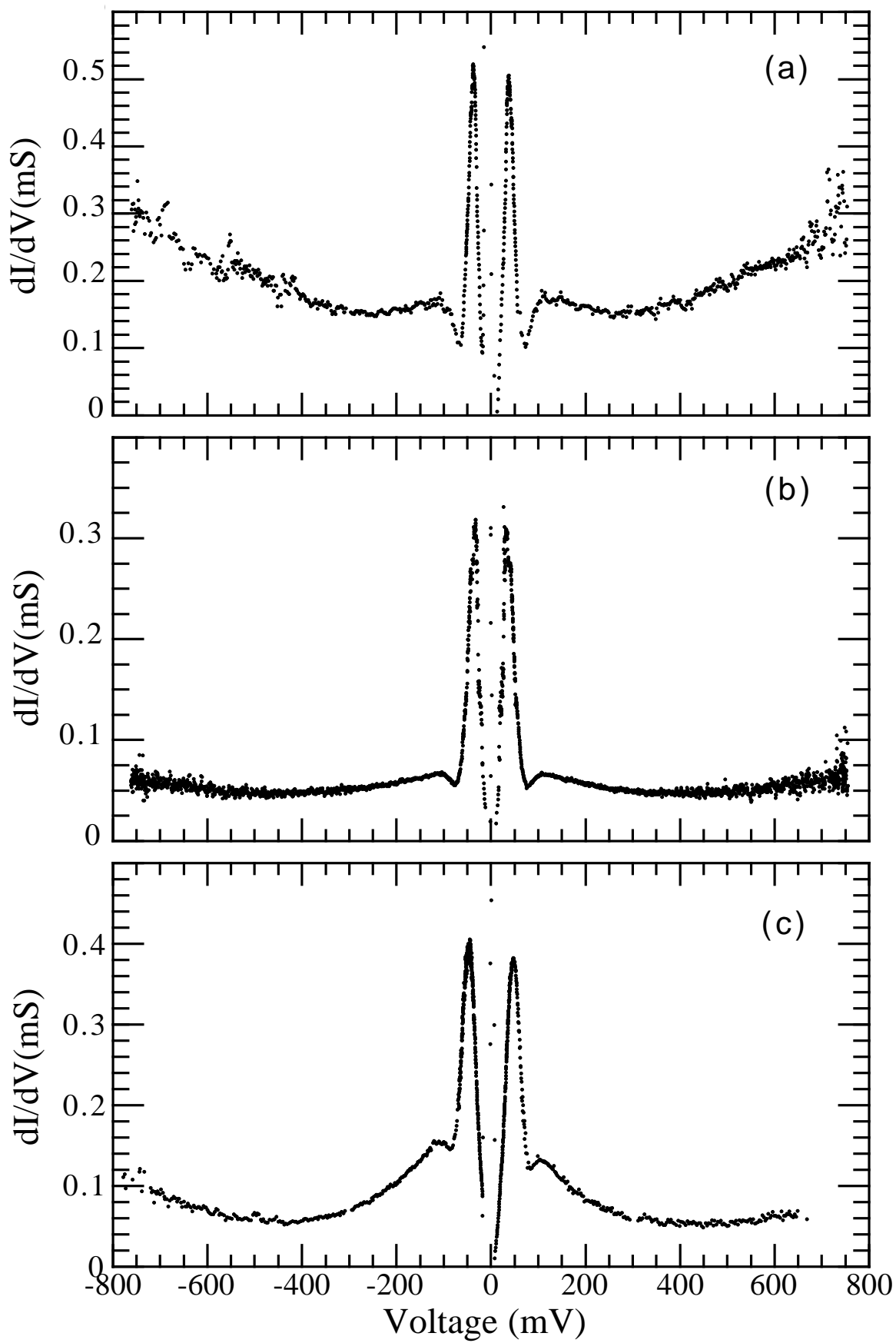
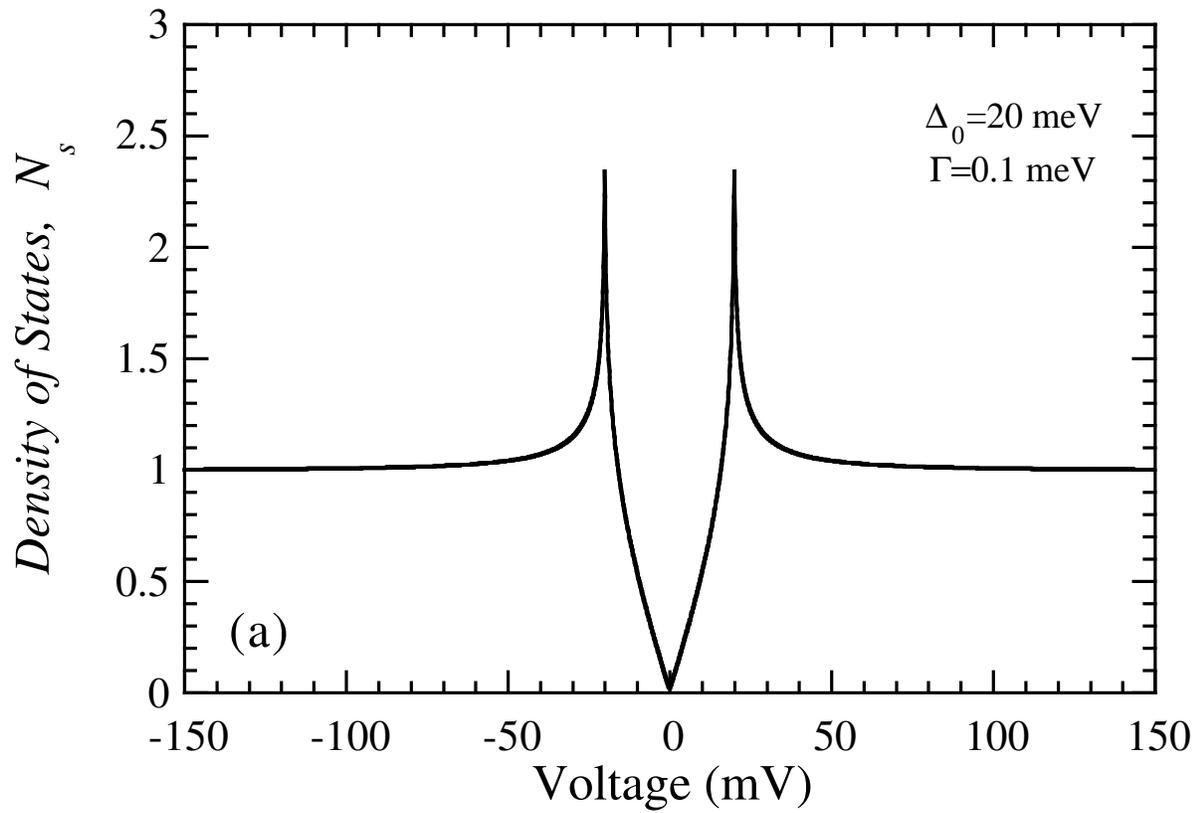
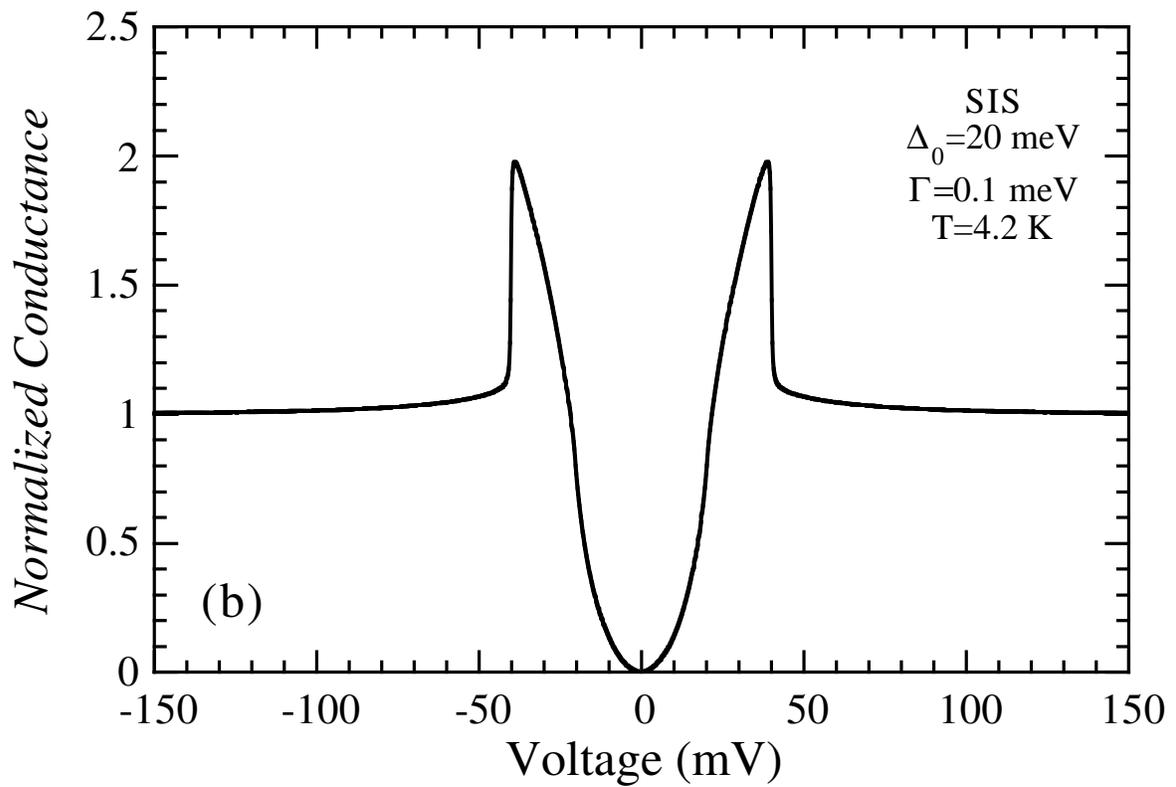
\end{document}